\pdfoutput=1
\RequirePackage{ifpdf}
\ifpdf 
\documentclass[pdftex]{sigma}
\else
\documentclass{sigma}
\fi

\usepackage{tabmac}

\numberwithin{equation}{section}

\begin{document}

\newcommand{\arXivNumber}{1503.09029}

\allowdisplaybreaks

\renewcommand{\thefootnote}{$\star$}

\renewcommand{\PaperNumber}{051}

\FirstPageHeading

\ShortArticleName{From Jack to Double Jack Polynomials via the Supersymmetric Bridge}

\ArticleName{From Jack to Double Jack Polynomials\\ via the Supersymmetric Bridge\footnote{This paper is a~contribution to the Special Issue on Exact Solvability and Symmetry Avatars
in honour of Luc Vinet.
The full collection is available at
\href{http://www.emis.de/journals/SIGMA/ESSA2014.html}{http://www.emis.de/journals/SIGMA/ESSA2014.html}}}

\Author{Luc LAPOINTE~$^\dag$ and Pierre MATHIEU~$^\ddag$}

\AuthorNameForHeading{L.~Lapointe and P.~Mathieu}

\Address{$^\dag$~Instituto de Matem\'atica y F\'{\i}sica, Universidad de
Talca, 2 norte 685, Talca, Chile}
\EmailD{\href{mailto:lapointe@inst-mat.utalca.cl}{lapointe@inst-mat.utalca.cl}}

\Address{$^\ddag$~D\'epartement de physique, de g\'enie physique et
d'optique, Universit\'e Laval,\\
\hphantom{$^\ddag$}~Qu\'ebec, Canada, G1V 0A6}
\Email{\href{mailto:pmathieu@phy.ulaval.ca}{pmathieu@phy.ulaval.ca}}

\ArticleDates{Received March 31, 2015, in f\/inal form June 25, 2015; Published online July 02, 2015}

 \Abstract{The Calogero--Sutherland model occurs in a large number of physical contexts, either directly or via its eigenfunctions, the Jack polynomials. The supersymmetric counterpart of this model, although much less ubiquitous, has an equally rich structure. In particular, its eigenfunctions, the Jack superpolynomials, appear to share the very same remarkable combinatorial and structural properties as their non-supersymmetric version. These super-functions are parametrized by superpartitions with f\/ixed bosonic and fermionic degrees.
Now, a truly amazing feature pops out when the fermionic degree is suf\/f\/iciently large:
the Jack superpolynomials {\it stabilize} and {\it factorize}. Their stability is with respect to their expansion in terms of an elementary basis where, in the stable sector, the expansion coef\/f\/icients become independent of the fermionic degree. Their factorization is seen when the fermionic variables are stripped of\/f in a suitable way which results in a product of two ordinary Jack polynomials
(somewhat modif\/ied by plethystic transformations), dubbed the double Jack polynomials.
 Here, in addition to spelling out these results, which were f\/irst obtained in the context of Macdonal superpolynomials, we provide a heuristic derivation of the Jack superpolynomial case by performing simple manipulations on the supersymmetric eigen-operators, rendering them independent of the number of particles and of the fermionic degree. In addition, we work out the expression of the Hamiltonian which characterizes the double Jacks. This Hamiltonian, which def\/ines a new integrable system, involves not only the expected Calogero--Sutherland pieces but also
combinations of the generators of an underlying af\/f\/ine ${\widehat{\mathfrak {sl}}_2}$ algebra.}


\Classification{05E05; 81Q60; 81R12; 37J35}

\renewcommand{\thefootnote}{\arabic{footnote}}
\setcounter{footnote}{0}

\section{Introduction}

The Calogero--Sutherland (CS) model is an integrable quantum many-body problem with a~surprisingly vast number of connections with dif\/ferent physical problems\footnote{For a representative sample of such connections and applications found in the period 1970--1995, see~\cite{DV}. A~short historical sketch of the f\/luctuations in the interest for these models, up to circa 2000, is presented in the introduction of~\cite{DLMnpb}.}.
Restricting ourself to the last 15 years, we can identify a number of important applications of the model itself or its eigenfunctions, the Jack polynomials. These are symmetric polynomials labeled by a partition $\lambda=(\lambda_1,\dots,\lambda_N)$ and depending upon a free parameter~$\alpha$ (which is the inverse of the CS coupling constant).

Early in the targeted period, a remarkable property of these polynomials has been unraveled by mathematicians \cite{FJMM}. For special partitions, namely with parts satisfying $ \lambda_i- \lambda_{i+k}\geq r$ and for $\alpha=-(k+1)/({r-1})$, the Jack polynomials vanish whenever $k+1$ of their variables are equal. This technical (dubbed clustering) property made them useful as trial wavefunctions for the fractional Hall ef\/fect~\cite{BH1}. In this context, the restriction on sequences of $k+1$ contiguous quasi-particle modes (the parts of the partition) is viewed as a generalized version of the Pauli exclusion principle.
In conformal f\/ield theory (CFT), these special Jacks correspond to the polynomial part of the correlators of $N$ generalized parafermionic f\/ields associated to the $W(k+1,k+r)$ minimal models~\cite{BGS,ES}.

A completely dif\/ferent connection with correlators in CFT is displayed in~\cite{Car,DC}.
Still in the context of CFT,
the remarkable connection between Virasoro singular vectors and Jack polynomials found in~\cite{AMOSa,AMOSb, MY} has been further clarif\/ied recently in~\cite{RW, SSAFR}. The technology of Jack polynomials can even be used to derive the spectrum of the Virasoro minimal mo\-dels~\mbox{\cite{RW, Wang}}. These applications have recently been lifted to the $\widehat{\mathfrak{sl}}(2)$ WZW model at fractional level~\cite{RWsl2}.

But perhaps the most spectacular application of the Jack polynomials in CFT is linked to the
Alday--Gaiotto--Tachikawa (AGT) correspondence~\cite{AGT}. This
 relates conformal blocks in two-dimensions CFT to the ${\rm U}(2)$ Nekrasov instanton partition function~\cite{Nek} pertaining to four-dimensional ${\rm SU}(2)$ supersymmetric Yang--Mills theory.
The Jack polynomials have been shown to be
 key components of a new AGT-motivated basis of states in 2d-CFT \cite{Alba}.
 More precisely, the Jack polynomials appear there in a generalized version which is indexed by a pair of partitions and decomposes into product of two Jacks with dif\/ferent arguments~\cite{Alba,MS}.

 Here we present a somewhat analogous type of generalization of the Jack polynomials also labelled by two partitions.
These new generalized Jacks arise directly from the construction of the supersymmetric counterparts of the Jack polynomials, the Jack superpolynomials~\cite{DLMcmp2}\footnote{We also use the terminology ``Jack polynomials in superspace''. For the motivation underlying this choice and the (non-)relationship of these objects with similarly-named polynomials discussed in the literature, see Appendix~\ref{appendixA}.}. The latter are eigenfunctions of the supersymmetric version of the CS model~\cite{SS}\footnote{The Jack superpolynomials also have
clustering properties similar to their non-supersymmetric counterparts~\cite{DLM0}. Moreover, they are related to the super-Virasoro singular vectors~\cite{ADM, DLMsvir}.}.
 It tuns out that for excited states with large fermionic degree, the eigenfunctions acquire an unexpected stability behavior.
More remarkably, in this stability sector, these eigenfunctions (after a minor transformation) factorize into a product of two Jack polynomials. This factorization is highly non-trivial: there is a sort of twisting in the coupling constant (the free parameter~$\alpha$), which is dif\/ferent for the two constituent Jacks, and a reorganization of the variables (technically: a plethystic transformation)\footnote{By contrast, the AGT-type double Jack polynomials~\cite{Alba,MS} are composed of two Jacks with dif\/ferent variables
(albeit corresponding to a less radical plethystic transformation), but the same coupling constant.}.
The factorized form of the eigenfunctions is referred to as the ``double Jack polynomials''.
We stress that the non-trivial structure of these double Jacks is inherited from the supersymmetric construction,
which thus serves as a bridge linking the Jacks to their double version.

 These peculiar properties of stability and factorization have f\/irst been observed at the level of the Macdonald generalization of the Jack superpolynomials \cite{BLM}. Here we make explicit the one-parameter limit characterizing the Jacks. However, instead of adapting to the Jack case the rather formal and technical argument of \cite{BLM}, we provide a simple heuristic argument leading to the stability and factorization properties. Basically, the strategy amounts to performing simple transformations of the two def\/ining eigen-operators of the supersymmetric CS model in order to make them independent of the number of variables and the fermionic degree.
 In addition, we unravel the integrable structure underlying the double-Jacks by constructing the Hamiltonian for which these are eigenfunctions\footnote{We note that the latter aspect would have been very dif\/f\/icult to study for the Macdonald case given the complexity of the supersymmetric form of the corresponding Ruisjenaars--Schneider model~\cite{BDM}.}.
Somewhat unexpectedly, this Hamiltonian is built in part from the generators of the nonnegative modes of an
${\widehat{\mathfrak s \mathfrak l}_2}$ algebra.

The article is organized as follows. In Section~\ref{section2}, we brief\/ly review the Calogero--Sutherland model and their eigenfunctions, emphasizing a Fock space representation to be used throughout the article. The supersymmetric CS model is introduced in Section~\ref{section3}, together with the Jack superpolynomials. For a suf\/f\/iciently high fermionic degree, the supersymmetric Hamiltonian eigenvalues are shown to be decomposable into two independent parts. This points toward the splitting of the Hamiltonian into two independent CS Hamiltonians and the corresponding factorization of its eigenfunctions. The resulting double Jack polynomials are def\/ined formally in Section~\ref{section4} and exemplif\/ied for simple cases, while their corresponding Hamiltonian is derived in Section~\ref{section5}.

\section{The Calogero--Sutherland model and Jack polynomials}\label{section2}

The CS model describes a system of $N$
 identical particles of mass $m$ lying on a circle of
circumference $L$ and interacting pairwise through the inverse of chord distance squared.
Setting $m=\hbar=1$ and $L=2\pi$, the Hamiltonian reads \cite{Suther1,Suther1+,Suther2}:
\begin{gather*} 
H^{\text{CS}}=\frac{1}{2}\sum_{i=1}^{N}p_i^2+\sum_{1\leq
i<j\leq N}\frac{\beta(\beta-1)}{4\sin^2 \tfrac12(x_i-x_j)},
\end{gather*}
 where $\beta$ is a
dimensionless real coupling constant and $[x_j,p_k]=i\delta_{jk}$.\footnote{See~\cite {KK} for an extensive and very clear presentation of the CS model.}
 To the ground state correspond the
following wavefunction and eigenvalue:
\begin{gather*} 
\psi_0(x)=\prod_{j<k}\big|\sin\tfrac12(x_j-x_k)\big|^\beta
\qquad \text{with}\quad E_0=
\frac{\beta^2 N(N^2-1)}{24}.
\end{gather*} It is
convenient to def\/ine
$z_j=e^{i x_j}$
and to factor out the contribution of the ground state
by redef\/ining a gauged Hamiltonian as $\psi_0^{-1}(H^{\text{CS}}
-E_0)\psi_0 /\beta$
and to set $\beta=1/\alpha$:
\begin{gather}\label{hjack}
{\mathcal H}^{(\alpha)} =\alpha \sum_{i=1}^N \left(z_i\partial_{z_i}\right)^2+\sum_{1 \leq i<j \leq N} \left(\frac{z_i+z_j}{z_i-z_j}\right)
\left(z_i\partial_{z_i}-z_j\partial_{z_j}\right).
\end{gather}
This is our starting point.

The symmetric and triangular eigenfunctions of \eqref{hjack} are known as
the
Jack polyno\-mials $J_\lambda^{(\alpha)}(z)$~\cite{LV}\footnote{For a
physicist introduction to the Jack polynomials, we refer to~\cite{For, KK}. A more mathematical presentation can be found in~\cite{Mac}.}, where the index $\lambda$ stands for a partition $\lambda=(\lambda_1,\lambda_2,\dots,\lambda_N)$, with the $\lambda_i$'s being non-negative integers such that $\lambda_i\geq \lambda_{i+1}$.
 Their
eigenvalues are
\begin{gather*}
 \varepsilon_\lambda^{(\alpha)}= 2\alpha n(\lambda') -2n(\lambda) + (N-1+\alpha)|\lambda|,
 \end{gather*}
 where \cite{Mac}
 \begin{gather}\label{defden}
{n(\lambda)= \sum_i (i-1) \lambda_i}=\sum_i\binom{\lambda_i'}{2}.
\end{gather}
Here $\lambda'$ is the conjugate of $\lambda$ obtained from $\lambda$ by replacing rows by columns in its diagrammatic representation, and $|\lambda|=\sum\limits_{i}\lambda_i$ is the degree of $\lambda$.
We will be interested in the behavior of the wavefunction when $N$ is large. It is thus preferable to remove the dependency in $N$ in the eigenvalue. For this, we note that $J_\lambda^{(\alpha)}(z)$ is homogeneous in the $z_i$'s, so that it is an eigenfunction of the momentum operator ${\mathcal P}$:
\begin{gather*} {\mathcal P} J^{(\alpha)}_\lambda(z)= \sum_{i} z_i \partial_{z_i} J_\lambda^{(\alpha)}(z)=|\lambda|J_\lambda^{(\alpha)}(z).
\end{gather*}
Our task is achieved by redef\/ining the Hamiltonian as
\begin{gather*} {\mathcal H}^{(\alpha)} \longrightarrow \hat {\mathcal H}^{(\alpha)}={\mathcal H}^{(\alpha)}- (N-1+\alpha)\sum_{i} z_i \partial_{z_i}.
\end{gather*}
Jack polynomials $J_\lambda^{(\alpha)}(z)$ are thus eigenfunctions of $\hat {\mathcal H}^{(\alpha)}$ with
eigenvalues
\begin{gather}\label{ephat}
\hat \varepsilon_\lambda^{(\alpha)}= 2\alpha n(\lambda') -2n(\lambda).
\end{gather}

In the large $N$ limit,
it is convenient to rewrite the Hamiltonian
in terms of power sums $ p_k= z_1^k+z_2^k+\cdots$. Since
$\hat {\mathcal H}^{(\alpha)}$ is a dif\/ferential operator of order two, it is suf\/f\/icient
to determine its action on the product~$p_m p_n$.
A direct
computation gives \cite{AMOSa, MP}
\begin{gather*}
\hat {\mathcal H}^{(\alpha)}=(\alpha-1)\sum_{n\geq 1} \big(n^2-n\big) p_n \partial_{p_n}
+\sum_{n,m\geq 1}[(m+n)p_m
p_n \partial_{p_{m+n}}+\alpha m n p_{m+n} \partial_{p_n} \partial_{p_m}].
\end{gather*}
This naturally leads to the Fock space representation
\begin{gather*}
\alpha
\hat {\mathcal H}^{(\alpha)}
= (\alpha-1) \sum_{\ell\geq 1} (\ell-1) a_\ell^\dagger a_\ell+ \sum_{k,\ell \geq 1}\big[ a_k^\dagger a_\ell^\dagger a_{k+\ell} +
a_{k+\ell}^\dagger a_k a_\ell \big],
\end{gather*}
where
\begin{gather*}
\big[a_k, a^\dagger_\ell\big] = k\alpha \delta_{k,\ell} \qquad {\rm and} \qquad [a_k, a_\ell]=\big[a_k^\dagger, a^\dagger_\ell\big]=0.
\end{gather*}
The correspondence with symmetric functions, together with $|0\rangle\longleftrightarrow 1$, is
\begin{gather}\label{amodes}
a^\dagger_k \longleftrightarrow p_k
\qquad {\rm and} \qquad
a_k \longleftrightarrow k \alpha {\partial_{p_k}}.
\end{gather}
This correspondence
 preserves the commutation relations.
In this representation, the eigenfunctions take the form of a combination of states
\begin{gather*}
J_{\lambda}^{(\alpha)}\big(a_1^\dagger,a_2^\dagger,a_3^\dagger,\dots\big) |0\rangle.
\end{gather*}
(To be clear, this mode expression is a formal representation of the Jack polynomial expanded in the power-sum basis through the correspondence~\eqref{amodes}.) For instance, up to a multiplicative
constant\footnote{In the monic normalization $J^{(\alpha)}_\lambda=m_\lambda+ \text{lower terms}$, where $m_\lambda$ is the monomial symmetric function, this coef\/f\/icient is $1/[2(1+\alpha)^2]$.}
\begin{gather*}
J^{(\alpha)}_{(3,1)}|0\rangle
\propto
\big[\big(a_1^\dagger\big)^4+(3\alpha-1)a_2^\dagger\big(a_1^\dagger\big)^2+2\alpha(\alpha-1)a_3^\dagger a_1^\dagger -\alpha\big(a_2^\dagger\big)^2-2\alpha^2a_4^\dagger\big] |0\rangle.
\end{gather*}

As a side remark, we point out that it is through the correspondence~\eqref{amodes}
 that the connection between Virasoro singular vectors and Jack polynomials is established \cite{AMOSa,AMOSb,MY,RW,SSAFR}.

\section{Supersymmetric version}\label{section3}

In order to supersymmetrize the CS model, we need to introduce anticommuting variables $\theta_1,\ldots,\theta_N$ and extend the CS Hamiltonian~$H$ in the following way:
\begin{gather*}
\hat{\mathcal H}^{(\alpha)}\to {\mathcal H}^{(\alpha)}_{{\rm susy}}
=\big\{Q,Q^\dagger\big\}=\hat{\mathcal H}^{(\alpha)}+\text{terms depending upon $\theta_i$},
\end{gather*}
for two fermionic charges $Q$ and $Q^\dagger$ of the form $Q=\sum\limits_i\theta_iA_i(x,p)$ and $Q^\dagger=\sum\limits_i\partial_{\theta_i}A^\dagger_i(x,p)$, where $A_i$ and $A_i^\dagger$ are f\/ixed by the requirement of reproducing the $\hat{\mathcal H}^{(\alpha)}$ term on the rhs of the above equation. This construction leads to
\begin{gather*}
{\mathcal H}^{(\alpha)}_{\text{{\rm susy}}}
= \hat {\mathcal H}^{(\alpha)} -2\sum_{1 \leq i<j \leq N} \frac{z_i z_j}{(z_i-z_j)^2}
(\theta_i-\theta_j) (\partial_{\theta_i} - \partial_{\theta_j}).
\end{gather*}
This operator is part of the tower of conserved quantities ${\mathcal H}_n$, $1\leq n\leq N$ (${\mathcal P}={\mathcal H}_1$ and ${\mathcal H}^{(\alpha)}_{\text{susy}} ={\mathcal H}_2$) that reduce to the usual (gauged) CS conservation laws in the absence of anticommuting variables.
But given that there are $2N$ degrees of freedom in the supersymmetric version, there are $N$ extra conserved charges that vanish when all $\theta_i=0$~\cite{DLMcmp2}. The f\/irst nontrivial representative of this second tower is
\begin{gather*}
{\mathcal I}^{(\alpha)}_{{\rm susy}}= \alpha \sum_{i=1}^N z_i \theta_i \partial_{z_i} \partial_{\theta_i}+
\sum_{1 \leq i < j \leq N} \frac{z_i \theta_j+z_j \theta_i}{z_i-z_j}(
\partial_{\theta_i}-\partial_{\theta_j}).
\end{gather*}

As a side remark, we mention that both expressions can be represented in the Fock space of a free boson, described by the modes $a_k$, $a_k^\dagger$ (with $k\geq 1$, i.e., without the zero mode) and a
free fermion, whose modes are denoted $b_k$, $b_k^\dagger$:\footnote{In a supersymmetric context, the modes of the partner free fermion should pertain to the Neveu--Schwarz sector, hence be half-integers. This can be achieved by redef\/ining $(b_k,b^{\dagger}_k)$ as $(b_{k+1/2},b^{\dagger}_{k+1/2})$ in the relation~\eqref{bvsp} below. However, this precision is not required in the present context.}
\begin{gather*}
\alpha{\mathcal H}^{(\alpha)}_{{\rm susy}} =
(\alpha-1) \sum_{\ell\geq 1}( \ell-1) a_\ell^\dagger a_\ell
+\sum_{k,\ell \geq 1}\big[ a_k^\dagger a_\ell^\dagger a_{k+\ell} +
a_{k+\ell}^\dagger a_k a_\ell \big]
 \\
\hphantom{\alpha{\mathcal H}^{(\alpha)}_{{\rm susy}} =}{} + \alpha(\alpha-1) \sum_{\ell\geq 1} \big(\ell^2 -\ell\big) b_\ell^\dagger b_\ell+\alpha \sum_{k,\ell \geq 1}
2\ell \big[ a_k^\dagger b_\ell^\dagger b_{k+\ell} + b_{k+\ell}^\dagger b_\ell a_k \big]
 \end{gather*}
and
\begin{gather*}
{\mathcal I}_{{\rm susy}}^{(\alpha)}= (\alpha -1) \sum_{\ell \geq 0} \ell b_\ell^\dagger b_\ell + \sum_{\ell \geq 0,\, k \geq 1} \big[ b_{\ell+k}^\dagger b_\ell a_k + a_{k} b_{\ell}^\dagger b_{\ell+k} \big].
\end{gather*}
The fermionic modes are governed by the anticommutation relations:
 \begin{gather*}
\big\{ b_k,b_\ell^\dagger \big\}= \delta_{k,\ell} \qquad \{ b_k,b_\ell \}
= \big\{ b_k^\dagger,b_\ell^\dagger \big\} =0 ,
\end{gather*}
and their correspondence with symmetric functions is
\begin{gather}\label{bvsp}
b_k^\dagger \longleftrightarrow \tilde p_k = \theta_1z_1^k+\theta_2 z_2^k +\cdots\qquad\text{and} \qquad b_k \longleftrightarrow \partial_{\tilde p_k}.
\end{gather}
Now, assuming a natural triangularity condition, the common eigenfunctions of ${\mathcal H}^{(\alpha)}_{{\rm susy}}$ and~${\mathcal I}^{(\alpha)}_{{\rm susy}}$ are the Jack polynomials in superspace, or Jack superpolynomials, denoted by~$J_\Lambda^{(\alpha)}(z,\theta)$~\cite{DLMcmp2}. They are homogeneous in $z$ and in $\theta$ and invariant under the exchange of pairs $(z_i,\theta_i) \longleftrightarrow
(z_j,\theta_j)$.
Their labelling index $\Lambda$ is a superpartition. Before displaying the eigenvalues, some notation related to superpartitions is required.

A superpartition $\Lambda$ is a pair of partitions
\begin{gather*}
\Lambda=(\Lambda^a;\Lambda^s)
\qquad \text{such that}\quad \begin{cases}
 \Lambda^s \ \text{is an ordinary partition},\\
\Lambda^a \ \text{is a partition with no repeated parts.}
\end{cases}
\end{gather*}
Note that the last part of~$\Lambda^a$ is allowed to be zero. We denote by $\Lambda^*$ the partition obtained by
reordering
in non-increasing order
the entries of~$\Lambda^a$ and~$\Lambda^s$ concatenated. The diagrammatic representation of $\Lambda$ is obtained by
putting dots at the end of the rows that come from $\Lambda^a$ (in such a way that dots never lie under an empty cell).
Here is an example:
\begin{gather*}
 \Lambda=( 4, 2, 0;3,2,1,1) \ \longleftrightarrow \
 ({ 4},3,{ 2},2,1,1,{ 0}) \ \longleftrightarrow \
\mbox{\tiny ${{{\tableau*[scY]{ & & & & \bl \bullet \cr & &
 \cr & & \bl \bullet \cr
&
\cr \cr \cr \bl \bullet }}}}$}
\end{gather*}
A superpartition is equally well described by the pair~$\Lambda^*$ and~$\Lambda^\circledast$, where the latter is the partition obtained by replacing dots by boxes, e.g., in the example above,
\begin{gather*}
\Lambda^{*}=\mbox{\tiny ${{{\tableau*[scY]{ & & & & \bl \cr & &
 \cr & & \bl \cr
& \cr \cr \cr \bl }}}}$}
 \qquad
\Lambda^{\circledast}=\mbox{\tiny ${{{\tableau*[scY]{ & & & & \cr & &
 \cr & & \cr
& \cr \cr \cr \cr }}}}$}
\end{gather*}
Finally, the bosonic degree of a superpartition is the number of boxes of $\Lambda^*$ and the fermionic degree, generally denoted by~$m$, is the number of dots in the diagram of $\Lambda$, that is, the number of parts of $\Lambda^a$.

We are now in position to give the eigenvalues of ${\mathcal H}^{(\alpha)}_{{\rm susy}}$ and ${\mathcal I}^{(\alpha)}_{{\rm susy}}$
corresponding to the eigenfunction
$J_{\Lambda}^{(\alpha)}$. These are respectively
\begin{gather*}
\varepsilon_{\Lambda}^{(\alpha)}=2 \alpha n({\Lambda^*}')-2n(\Lambda^*)
 \qquad {\text{and}}\qquad e_\Lambda^{(\alpha)}= \alpha |\Lambda^a|- \big|{\Lambda'}^{a}\big|.
\end{gather*}

In the supersymmetric case, we are not only interested in the large $N$ limit but also in the large $m$ limit (actually, in the large $m$ and $N-m$ limits).
We thus want to extract from the above two eigenvalues, their dependence on $m$ which is somewhat hidden. For this, we f\/irst notice that when $m$ is large (relative to the size of~$\Lambda$, an estimation that is made precise in~\eqref{bornem}), there are circles in every possible positions
in the diagram of $\Lambda$.\footnote{Here is a technical precision that could safely be skipped. There are unimportant exceptions to the statement that when $m$ is suf\/f\/iciently large (meaning larger or equal to its lower bound, which is~$|\lambda|+|\mu|$ for~$\lambda$ and~$\mu$ def\/ined below), there are dots in every possible positions. That all allowed slots are f\/illed by dots is true when $m\geq \ell(\lambda)+1+\mu_1$, $\ell(\lambda)$ being the length of the partition $\lambda$. Since $|\lambda|+|\mu|+1 \geq \ell(\lambda)+1+\mu_1$, the statement is always true for instance when $m \geq |\lambda|+|\mu|+1$.}
 As such, the circles can be ignored and we observe that
$\Lambda^*$ dif\/fers slightly from its core $\delta^{(m)}=(m-1,m-2,\dots,1,0)$. In the diagrammatic representation of $\Lambda^*$, the deviations to the core are located at the top right and at the bottom left of the diagram. We thus see that the superpartition can be disentangled into its fermionic core plus two small partitions $\lambda$ and $\mu$ such that $\Lambda=(\lambda+\delta^{(m)};\mu)$ \cite{BLM}.\footnote{For the $+$ operation, the parts add up. For example, we have $(3,1)+(4,2,2)=(7,3,2)$.} For instance,
 for $m=8$, we have
\begin{gather*}
\Lambda= \ \mbox{\tiny ${{{\tableau*[scY]{ & & & & & & & & & \bl \bullet
 \cr & & & & & & & \bl \bullet
 \cr & & & & & \bl \bullet
 \cr & & & & \bl \bullet
\cr
& & & \bl \bullet
\cr & &
 \cr & & \bl \bullet
\cr & \bl \bullet \cr \cr \bl \bullet }}}}$}
\quad \longleftrightarrow \quad
\Lambda= \ \mbox{\tiny ${{{\tableau*[scY]{ & & & & & & & \fl & \fl & \bl \phantom{\bullet}
 \cr & & & & & & \fl & \bl \phantom{\bullet}
\cr & & & & & \bl \phantom{\bullet}
 \cr & & & & \bl \phantom{\bullet}
\cr
& & & \bl \phantom{\bullet}
\cr & & \tf
 \cr & \tf & \bl \phantom{\bullet}
\cr \tf & \bl \phantom{\bullet} \cr \tf \cr \bl \phantom{\bullet} }}}}$}
\quad \longleftrightarrow \quad
\mbox{\tiny ${{{\tableau*[scY]{ & & & & & & & \bl & \bl & \fl
& \fl & \bl &\bl = \lambda
 \cr & & & & & & \bl & \bl & \bl & \fl
\cr & & & &
\cr & & & \cr
& & \cr &
\cr \cr \bl \cr \bl \cr \tf & \tf & \tf & \bl &\bl = \mu
\cr \tf }}}}$}
\end{gather*}
 It is clear that~$\Lambda$ is fully characterized by $m$ and the pair
$ (\lambda,\mu) $ (whose total degree is much less than that of~$\Lambda$).
 The main advantage of this diagrammatic decomposition is that it implies readily that when~$m$ is large the conjugate of~$\Lambda$ is $\Lambda'=(\mu'+\delta^{(m)};\lambda')$.

Let us reformulate the eigenvalues in terms of the data~$\lambda$, $\mu$ and~$m$.
For the ${\mathcal I}^{(\alpha)}_{{\rm susy}}$ eigenvalue, the computation is easy and yields
\begin{gather*}
e_{\Lambda}^{(\alpha)} \,\overset{m\,\text{large}}\longrightarrow\, e_{\lambda,\mu}^{(\alpha)}= \alpha |\lambda|- |\mu|+(\alpha-1)m(m-1)/2.
\end{gather*}
We can easily remove the dependency in $m$ in the eigenvalue by redef\/ining ${\mathcal I}^{(\alpha)}_{{\rm susy}} $ as follows:
\begin{gather*}
{\mathcal I}^{(\alpha)}_{{\rm susy}} \longrightarrow \hat {\mathcal I}^{(\alpha)} _{{\rm susy}}= {\mathcal I}^{(\alpha)} _{{\rm susy}}- (\alpha-1) {\mathcal M} ({\mathcal M}-1)/2, \qquad\text{where}
\quad {\mathcal M}= \sum_i \theta_i \partial_{\theta_i}.
\end{gather*}
This subtraction is well def\/ined since
${\mathcal M}$ is also a conserved quantity. The modif\/ied eigenvalue reads then
\begin{gather}\label{ehat}
\hat e_{\lambda,\mu}^{(\alpha)}= \alpha |\lambda|- |\mu|.
\end{gather}

The eigenvalues of ${\mathcal H}^{(\alpha)}_{{\rm susy}}$ can also be reformulated in terms of $\lambda$, $\mu$ and $m$, again keeping in mind that this is valid only for suf\/f\/iciently large~$m$. Observe that\footnote{For the $\cup$ operation, the rows of the second partition are inserted into the f\/irst one; for instance $(3,1)\cup(4,2,2)=(4,3,2,2,1)$.}
\begin{gather*}
\Lambda^*=\big(\lambda+\delta^{(m)}\big)\cup \mu\quad \implies \quad \Lambda^*_i\in\big\{\lambda_j+m-j\,|\,1\leq j\leq m\big\}\cup\big\{ \mu_k\,|\,1\leq k\leq \ell(\mu)\big\},
\end{gather*}
and similarly for ${\Lambda^*}'=(\mu'+\delta^{(m)})\cup \lambda'$,
where $\ell(\mu)$ is the length of the partition $\mu$ (the number of non-zero parts).
Fortunately, the calculation of $n(\Lambda^*)$ (and $n({\Lambda^*}')$) is independent of the precise relationship between the indices $i$ and $j$, $k$ in the above notation
if we use the second expression of~$n(\lambda)$ given in~\eqref{defden}.
Let us f\/irst consider
\begin{gather*}
n({\Lambda^*}') =\sum_i\binom{\Lambda_i^*}{2}=\sum_{i=1}^m\binom{\lambda_i+m-i}{2}+\sum_{i=1}^{\ell(\mu)}\binom{\mu_i}{2} \\
\hphantom{n({\Lambda^*}')}{} =\sum_{i=1}^m\left[\binom{\lambda_i}{2}+\binom{m-i}{2}+\lambda_i(m-i)\right]+n(\mu') \\
\hphantom{n({\Lambda^*}')}{} =n(\lambda')+n(\mu')+\sum_{i=1}^{\ell(\lambda)}\lambda_i[(m-1)-(i-1)]+m(m-1)(m-2)/6 \\
\hphantom{n({\Lambda^*}')}{} =n(\lambda')+n(\mu')+(m-1)|\lambda|-n(\lambda)+m(m-1)(m-2)/6,
\end{gather*}
where in the last step, we use the f\/irst expression in~\eqref{defden}. For the computation of~$n(\Lambda^*)$, we simply replace $\lambda$ and $\mu$ by $\mu'$ and $\lambda'$
respectively in the previous expression to get:
\begin{gather*}
n(\Lambda^*)=n(\mu)+n(\lambda)+(m-1)|\mu|-n(\mu')+m(m-1)(m-2)/6 .
\end{gather*}
Combining these two expressions yields
\begin{gather*}
\varepsilon_{\Lambda}^{(\alpha)} \,\overset{m\,\text{large}}\longrightarrow\, \varepsilon_{\lambda,\mu}^{(\alpha)}= (\alpha+1)
\hat\varepsilon_\lambda^{ (\alpha/(\alpha+1) )}+\hat\varepsilon_\mu^{(\alpha+1)}+
2 (m-1)
\hat e_{\lambda,\mu}^{(\alpha)} +(\alpha-1)m(m-1)(m-2)/3,
\end{gather*}
where $\hat\varepsilon_\mu^{(\alpha)}$ is def\/ined in \eqref{ephat}.
 We can thus remove the dependency in $m$ in the eigenvalue by redef\/ining $ {\mathcal H}^{(\alpha)}_{{\rm susy}}$ as
\begin{gather*}
 {\mathcal H}^{(\alpha)}_{{\rm susy}} \longrightarrow\hat {\mathcal H}^{(\alpha)}_{{\rm susy}} = {\mathcal H}^{(\alpha)}_{{\rm susy}}- {2}({\mathcal M}-1) \hat {\mathcal I}^{(\alpha)}_{{\rm susy}} -(\alpha-1){\mathcal M}({\mathcal M}-1)({\mathcal M}-2)/3.
 \end{gather*}
The $\hat{\mathcal H}^{(\alpha)}_{{\rm susy}}$ eigenvalue is then simply
\begin{gather} \label{eigensum}
\hat\varepsilon_{\lambda,\mu}^{(\alpha)}=(\alpha+1)
\hat\varepsilon_\lambda^{ (\alpha/(\alpha+1) )}+ \hat\varepsilon_\mu^{(\alpha+1)}.
\end{gather}

The two expressions \eqref{ehat} and \eqref{eigensum} are the results we were looking for, the conclusion of our heuristic argument based on simple transformations of the two def\/ining
Jack-superpolynomials' eigen-operators. On the one hand, that the $m$-dependence of the eigenvalues can be removed is an indication of the stability property of the eigenfunctions. On the other hand, the decoupling of these two eigenvalues
into two independent sectors $\lambda$ and $\mu$ suggests the factorization property. More precisely, the decomposition of
$\hat\varepsilon_{\lambda,\mu}^{(\alpha)}$ as the sum of $\hat\varepsilon_{\lambda}^{ (\alpha/(\alpha+1) )} $ and $\hat\varepsilon_{\mu}^{(\alpha+1)}$ hints that, in the large $m$ limit, the eigenfunction $J_\Lambda^{(\alpha)}$ should somehow factorize into a product of the form $J_\lambda^{ (\alpha/(\alpha+1) )}$ times $J_\mu^{(\alpha+1)}$ (with \eqref{ehat} being compatible with the respective modif\/ications of the coupling constants).

These two conclusions are indeed conf\/irmed:
the eigenfunctions both stabilize and factorize (after a certain transformation that will be explained
in equation~\eqref{map})\footnote{When the Jack superpolynomial is expressed in terms of the variables $(x,\theta)$ rather than in
modes, the transformation is simply
\begin{gather*}
\Delta_m(x)^{-1}\partial_{\theta_m}\cdots \partial_{\theta_1} J_\Lambda^{(\alpha)}(x,\theta), \qquad \text{where}\quad \Delta_m(x)=\prod_{1\leq i<j\leq m}(x_i-x_j).
\end{gather*}} for~\cite{BLM}:
\begin{gather} \boxed{m\geq|\lambda|+|\mu|}.
\label{bornem}
\end{gather}

Let us consider a simple example.
For $(\lambda,\mu)= ((1),(1))$
the $m=1,2,3,4$ eigenfunctions read respectively
\begin{gather*}
 \bigl[ { b_0^\dagger} \big(a_1^\dagger\big)^2+\alpha {b_1^\dagger} a_1^\dagger -{b_0^\dagger} a_2^\dagger -\alpha
 {b_2^\dagger } \bigr] | 0 \rangle, \\
 \bigl[ {b_1^\dagger b_0^\dagger} \big(a_1^\dagger\big)^2+(\alpha-1) {b_2^\dagger b_0^\dagger} a_1^\dagger -\alpha {b_2^\dagger b_1^\dagger} -\alpha
 {b_3^\dagger b_0^\dagger} \bigr] | 0 \rangle,\\
 \bigl[ {b_2^\dagger b_1^\dagger b_0^\dagger} \big(a_1^\dagger\big)^2+(\alpha-1) {b_3^\dagger b_1^\dagger b_0^\dagger} a_1^\dagger
-\alpha {b_3^\dagger b_2^\dagger b_0^\dagger } -\alpha
 {b_4^\dagger b_1^\dagger b_0^\dagger} \bigr] | 0 \rangle, \\
 \bigl[ {b_3^\dagger b_2^\dagger b_1^\dagger b_0^\dagger} \big(a_1^\dagger\big)^2+(\alpha-1) {b_4^\dagger b_2^\dagger b_1^\dagger b_0^\dagger} a_1^\dagger
-\alpha {b_4b_2^\dagger b_1^\dagger b_0^\dagger } -\alpha
 {b_5^\dagger b_2^\dagger b_1^\dagger b_0^\dagger} \bigr] | 0 \rangle .
\end{gather*}
Clearly, the $m=1$ ($<|\lambda|+|\mu|=2$) wavefunction does not belong to the stable sector. For $m\geq 2$, the coef\/f\/icients and the $a^\dagger$ content of each term are always the same. This is the stability property.

Although they stabilize, the
eigenfunctions still depend on $m$.
However, consider the map
\begin{gather}\label{map}
\text{for}\ \big(\Lambda^a;\Lambda^s\big) \longleftrightarrow
(\lambda,\mu)\colon \quad b^\dagger_{\Lambda^a} a^\dagger_{\Lambda^s} |0\rangle \longleftrightarrow s_\lambda(y) p_{\mu}(y,z),
\end{gather}
where
\begin{gather*}
p_\mu=p_{\mu_1}\cdots p_{\mu_\ell}\qquad \text{with}\quad
 p_n(y,z)=\sum_{i=1}^m y_i^n + \sum_{i=1}^{N-m}z_i^n,
\end{gather*}
and $s_\lambda$ is the Schur function. Observe that $p_n(y,z)$ is simply $p_n$ in the variables
$y_1,y_2,\dots,y_m, z_1$, $z_2,\dots, z_{N-m}$.
This maps the above eigenfunctions corresponding to the values $m=2,3,4$ to (inserting the proper normalization){\samepage
\begin{gather*}
J_{(1),(1) }^{(\alpha)}(y,z)=\frac1{(1+\alpha)}\big[p_1(y,z)^2 + (\alpha-1) s_1(y)p_1(y,z)-\alpha s_{1,1}(y) -\alpha s_2(y)\big].
\end{gather*}
The stability has now been lifted to the full structure of the eigenfunction.}

But in addition, the map~\eqref{map} captures the factorization property suggested by the form of the eigenvalues. Using the Pieri rule for Schur functions \cite{Ber,Mac} to express the
sum of the last two terms in a product form,
\begin{gather*}
s_{1,1}(y)+s_2(y)=s_1(y) s_1(y),
\end{gather*}
we see that
$J_{(1),(1) }^{(\alpha)}$ can also be written in a product form
\begin{gather}\label{exsimple}
J_{(1),(1) }^{(\alpha)}(y,z)= \frac1{(1+\alpha)} {\bigl( p_1(y,z)+\alpha s_1(y) \bigr)}
 {\bigl(p_1(y,z)-s_1(y)\bigr)}.
\end{gather}
This is a simple illustration of the announced factorization.

\section{Double Jack polynomials}\label{section4}

In general, the action of the map \eqref{map} at the level of Jack polynomials is \cite{BLM}
 \begin{gather*} J^{(\alpha)}_\Lambda \big(b_1^\dagger,b_2^\dagger,\dots, a_1^\dagger,a_2^\dagger,\dots\big)|0\rangle\longleftrightarrow
J_{\lambda,\mu}^{(\alpha)}(y,z),
\end{gather*}
 where
 \begin{gather}\label{Jdouble}
 J_{\lambda,\mu}^{(\alpha)}(y,z)= J_\lambda^{(\alpha/(\alpha+1))}\left[Y+\frac{1}{\alpha+1}Z \right]
 J_\mu^{(\alpha+1)}(z).
\end{gather}
 Here
 we use the plethystic notation (see, e.g.,~\cite{Ber,Las}). In our case, it simply means
that if $J^{(\alpha/(\alpha+1))}_\lambda(p_1,p_2,p_3,\dots)$ is the expression of $J^{(\alpha/(\alpha+1))}_\lambda$ in terms of power-sums, then
 \begin{gather*}
J^{(\alpha/(\alpha+1))}_\lambda\left[Y+\frac{1}{\alpha+1}Z\right] \\
\qquad{} = J^{(\alpha/(\alpha+1))}_\lambda\left(p_1(y)+\frac{1}{\alpha+1}p_1(z), p_2(y)+\frac{1}{\alpha+1}p_2(z), p_3(y)+\frac{1}{\alpha+1}p_3(z),\dots\right)
\end{gather*}
that is, $J^{(\alpha/(\alpha+1))}_\lambda[Y+Z/(1+\alpha)]$ is obtained from the expansion of $J^{(\alpha/(\alpha+1))}_\lambda(z)$ in terms of power-sums
by replacing $p_n$ by $p_n(y)+\frac{1}{\alpha+1}p_n(z)$.

Let us recover \eqref{exsimple} from the general expression~\eqref{Jdouble}. This is a particularly simple case given that $J_{(1)}^{(\alpha)}=s_{(1)}=m_{(1)}=p_{1}$. With $p_{1}(y,z)=p_{1}(y)+p_{1}(z)$,
$J_{(1),(1) }^{(\alpha)}(y,z)$ becomes
\begin{gather*}
J_{(1),(1) }^{(\alpha)}(y,z)=\left(p_1(y)+\frac1{\alpha+1}p_1(z)\right) p_1(z),
\end{gather*}
which is indeed of the form~\eqref{Jdouble}.
Here is a slightly more complicated example:
\begin{gather*}
J^{(\alpha)}_{(2),(0)}(y,z)=\frac1{(1+\alpha)(1+2\alpha)}\big[ p_1(y,z)^2+\alpha p_2(y,z)+ 2 \alpha s_{(1)}(y) p_1(y,z)+2\alpha^2 s_{(2)}(y)\big].
\end{gather*}
With $s_{(2)}=(p_1^2+p_2)/2$, simple algebra yields
\begin{gather*}
J^{(\alpha)}_{(2),(0)}(y,z)=\frac{(1+\alpha)}{(1+2\alpha)}\left(p_1[X]^2+\frac{\alpha}{\alpha+1}p_2[X]\right)= J^{(\alpha/(\alpha+1))}_{(2)}[X]
\end{gather*}
with $X=Y\!+\!Z/(\alpha\!+\!1)$ and where in the last step, we used the expression $J^{(\alpha)}_{(2)}\!=\!(p_1^2\!+\alpha p_2)/(1+\alpha)$.

A more formal
 characterization of $ J_{\lambda,\mu}^{(\alpha)}(y,z)$, which we call the {\it double Jack polynomials}, is as follows~\cite{BLM}. They are the
unique bi-symmetric functions such that
 \begin{gather*}
J_{\lambda,\mu}^{(\alpha)}(y,z) = s_{\lambda}(y) s_{\mu}(z) + {\rm smaller~terms}
\end{gather*}
 {and}	
\begin{gather*}
\bigl \langle J_{\lambda,\mu}^{(\alpha)}(y,z), J_{\nu,\kappa}^{(\alpha)}(y,z)
\bigr\rangle = 0
\qquad{\rm if} \quad (\lambda,\mu) \neq (\nu,\kappa).
\end{gather*}
The triangularity condition that specif\/ies the ``smaller terms'' refers to the double version of the dominance ordering:
\begin{gather*}
(\lambda,\mu) \geq (\nu,\kappa) \ \iff \ |\lambda|+|\mu|=|\nu|+|\kappa| , \\
\sum_{i=1}^\ell (\lambda_i -\nu_i) \geq 0 , \qquad 
|\lambda|-|\nu|+
\sum_{j=1}^\ell (\mu_j -\kappa_j) \geq 0 \qquad \forall \, \ell ,
\end{gather*}
while the orthogonality condition refers to the scalar product
\begin{gather} \label{scalard}
\bigl \langle s_\lambda(y) p_{\mu}(y,z), s_\nu(y) p_{\kappa}(y,z) \bigr \rangle
=\delta_{\lambda \nu} \delta_{\mu \kappa} z_{\mu} \alpha^{\ell(\mu)}
\end{gather}
with $z_\mu=\prod\limits_{i\geq 1}i^{n_\mu(i)} n_\mu(i)!$, $n_\mu(i)$ being the multiplicity of the part~$i$ in~$\mu$. Observe that this scalar product has the form
\begin{gather*}
\bigl \langle \bullet,\bullet \bigr \rangle =
\bigl \langle \cdot,\cdot \bigr \rangle _{\rm Schur}^y
\bigl \langle \cdot,\cdot \bigr \rangle _{\rm Jack}^{y,z} ,
\end{gather*}
 where $\bigl \langle \cdot,\cdot \bigr \rangle _{\rm Schur}^y$ is the scalar product with respect to which the Schur functions
$s_\lambda(y)$ are orthonormal while $\bigl \langle \cdot,\cdot \bigr \rangle _{\rm Jack}^{y,z} $ is the scalar product
with respect to which the Jack polynomials $J_\lambda^{(\alpha)}(y,z)$ are orthogonal ($J_\lambda^{(\alpha)}(y,z)$ being the usual Jack polynomials in the variables
$y_1,y_2,\dots$, $y_m,z_1,z_2,\dots,z_{N-m}$).

\section[The double CS model and an emerging ${\widehat{\mathfrak s \mathfrak l}_2}$]{The double CS model and an emerging $\boldsymbol{{\widehat{\mathfrak s \mathfrak l}_2}}$}\label{section5}

Let us now unravel the integrable model whose eigenfunctions are the double Jack polynomials. The factorized expression~\eqref{Jdouble} of these polynomials and the splitting of the eigenvalue displayed in~\eqref{eigensum} readily indicate that the underlying Hamiltonian~${\mathcal H}_D$ is a sum of two CS Hamiltonians, albeit with modif\/ied coupling constants and involving unusual variables:
\begin{gather*}
{\mathcal H}_D = (\alpha+1) {\mathcal H}_1 + {\mathcal H}_2,
\end{gather*}
where
\begin{gather*}\mathcal H_1=\hat {\mathcal H}^{(\alpha/(\alpha+1))}\qquad\text{with}\quad
 \begin{cases}
 p_n \mapsto p_n[X],
 \\ \partial_{p_n} \mapsto \partial_{p_n[X]},
 \end{cases}\qquad\text{where}\quad X=Y+(\alpha+1)^{-1}Z,
\end{gather*}
and
\begin{gather*}\mathcal H_2=\hat {\mathcal H}^{(\alpha+1)}
\qquad\text{with}\quad\begin{cases} p_n\mapsto p_n(z), \\ \partial_{p_n} \mapsto \partial_{p_n(z)}.\end{cases}
\end{gather*}
Note that ${p_n[X]}$ and ${p_n(z)}$ are considered to be independent. Being the sum of two independent integrable Hamiltonians, ${\mathcal H}_D$ trivially characterizes a new integrable model.

However, the above splitting of ${\mathcal H}_D$ is not very interesting since it is hard to give a physical meaning to the power-sums ${p_n[X]}$, ${p_n(z)}$
and their derivatives.
The structure of the scalar product \eqref{scalard} points toward a more interesting choice of variables, namely $p_n(y)$ and $p_n(y,z)$, whose adjoints are $n \partial_{p_n(y)}$ and $n\alpha \partial_{p_n(y,z)}$ respectively.
With
\begin{gather*}
 X= Y+\frac1{\alpha+1}Z=\frac{\alpha}{\alpha+1}Y+\frac1{\alpha+1}(Y+Z),
\end{gather*}
the
change of variables is thus
\begin{gather}
 p_n[X]= \frac{\alpha}{\alpha+1}p_n(y) + \frac{1}{\alpha+1}p_n(y,z), \qquad
 p_n(z)= p_n(y,z)-p_n(y) ,\label{lesp}
\end{gather}
which gives (using the chain rule in two variables)
\begin{gather}
 \partial_{p_n[X]}=\partial_{p_n(y)}+ \partial_{p_n(y,z)},\qquad
 \partial_{p_n(z)}= \frac{\alpha}{\alpha+1}\partial_{p_n(y,z)} - \frac{1}{\alpha+1}\partial_{p_n(y)}.\label{lesdp}
\end{gather}
These expressions are readily checked by verifying that they satisfy the commutation relations:
\begin{alignat*}{3}
 & \bigl[\partial_{p_n[X]},{p_m[X]}\bigr]=\delta_{n,m} ,
\qquad && [\partial_{p_n(z)},p_m(z)]=\delta_{n,m} ,& \\
 & [\partial_{p_n[X]},{p_m(z)}]=0, \qquad && 
 \bigl[\partial_{p_n(z)},{p_m[X]}\bigr]=0 .&
\end{alignat*}
For these manipulations, we stress that $p_n(y)$ and $p_n(y,z)$ are considered to be independent, meaning:
\begin{gather*}
[\partial_{p_n(y)},p_m(y,z)]=[\partial_{p_n(y,z)},p_m(y)]=0.
\end{gather*}
We then substitute \eqref{lesp} and \eqref{lesdp} into $(\alpha+1) \mathcal H_1 + \mathcal H_2$.
The result, obtained after
 straightforward manipulations, is best rewritten in terms of two independent sets of bosonic modes def\/ined as
\begin{gather*}
A_n^\dagger=p_n(y)\qquad \text{and}\qquad A_n=n \partial_{p_n(y)}
\qquad \big(\Rightarrow \ \big[A_k, A^\dagger_\ell\big] = k\delta_{k,\ell} \big),
\end{gather*}
together with
\begin{gather*}
 a_n^\dagger=p_n(y,z)\qquad \text{and}\qquad a_n=n \alpha \partial_{p_n(y,z)}
\qquad \big(\Rightarrow \ \big[a_k, a^\dagger_\ell\big] = k {\alpha} \delta_{k,\ell} \big).
\end{gather*}
The resulting form of ${\mathcal H}_D$ is
\begin{gather}
 \alpha \mathcal H_{D} = \sum_{k,\ell \geq 1}\big[ a_k^\dagger a_\ell^\dagger a_{k+\ell} +
a_{k+\ell}^\dagger a_k a_\ell \big]
+ (\alpha-1) \sum_{\ell\geq 1} (\ell-1) a_\ell^\dagger a_\ell - \alpha \sum_{\ell\geq 1} (\ell-1) \big[a_\ell^\dagger A_\ell +
A_\ell^\dagger a_\ell \big] \nonumber \\
\hphantom{\alpha \mathcal H_{D} =}{} +\alpha \sum_{k,\ell \geq 1}\big[2 a_k^\dagger A_\ell^\dagger A_{k+\ell} +
a_{k+\ell}^\dagger A_k A_\ell \big] + \alpha \sum_{k,\ell \geq 1}\big[ A_k^\dagger A_\ell^\dagger a_{k+\ell} +
2 A_{k+\ell}^\dagger A_k a_\ell \big] \nonumber \\
\hphantom{\alpha \mathcal H_{D} =}{}
 + \alpha (\alpha-1) \sum_{k,\ell \geq 1}\big[ A_k^\dagger A_\ell^\dagger A_{k+\ell} +A_{k+\ell}^\dagger A_k A_\ell \big] .\label{HD}
\end{gather} It turns out that $\mathcal H_D$ can be reexpressed as
\begin{gather*}
 {\mathcal H}_D = \hat {\mathcal H}^{(\alpha)}_{y,z} + (\alpha-1) \hat {\mathcal H}^{(1)}_{y} + \big[Q_1,\hat {\mathcal H}^{(1)}_{y}\big]
-\frac{1}{2}[Q_1,Q_2],
\end{gather*}
where
\begin{gather*}
Q_1= \sum_{\ell\geq 1} \frac{1}{\ell} \big[ a_\ell^\dagger A_\ell-
A_\ell^\dagger a_\ell \big] \qquad {\rm and} \qquad Q_2= \sum_{\ell\geq 1} ({\ell-1}) \big[ A_\ell^\dagger A_\ell -\frac{1}{\alpha}
a_\ell^\dagger a_\ell \big] .
\end{gather*}
Note that $\hat {\mathcal H}^{(\alpha)}_{y,z}$ is $\hat {\mathcal H}^{(\alpha)}$ in the variables $y_1,y_2,\dots,y_m,z_1,z_2,\dots,z_{N-m}$,
so that
\begin{gather*} \alpha \hat {\mathcal H}^{(\alpha)}_{y,z}=\sum_{k,\ell \geq 1}\big[ a_k^\dagger a_\ell^\dagger a_{k+\ell} +
a_{k+\ell}^\dagger a_k a_\ell \big]
+ (\alpha-1) \sum_{\ell\geq 1} (\ell-1) a_\ell^\dagger a_\ell.
\end{gather*}
Similarly, $\hat {\mathcal H}^{(1)}_{y}$ is $\hat {\mathcal H}^{(\alpha)}$ in the variables $y_1,y_2,\dots,y_m$ but evaluated at $\alpha=1$:
\begin{gather*}
\alpha(\alpha-1)\hat {\mathcal H}^{(1)}_{y}=\alpha (\alpha-1) \sum_{k,\ell \geq 1}\big[ A_k^\dagger A_\ell^\dagger A_{k+\ell} +A_{k+\ell}^\dagger A_k A_\ell \big] .
\end{gather*}
Next, it is simple to check that $\alpha [Q_1,\hat {\mathcal H}^{(1)}_{y}]$ yields the second line in \eqref{HD}.
Therefore, parts of the constituents of ${\mathcal H}_D$ have a direct interpretation in terms of variables. However, this is not the case for~$Q_1$ and~$Q_2$.
Note that the action of $Q_1$ amounts to
exchanging the~$a$ and~$A$ modes (which thereby appears to be a remnant of the action of a supersymmetric charge).
Nevertheless,
it turns out that $Q_1$ and $Q_2$ have a nice Lie algebraic interpretation. More precisely, both are combinations of
the generators of an underlying af\/f\/ine ${\widehat{\mathfrak s \mathfrak l}_2}$ algebra (whose existence is not surprising
in the presence of two independent inf\/inite sets of bosonic modes). It is straightforward to verify that the operators
 \begin{gather*}
e^{(k)}= \frac{1}{\sqrt{\alpha}}\sum_{\ell \geq 1} \ell^{k-1} A_\ell^\dagger a_\ell , \qquad f^{(k)}= \frac{1}{\sqrt{\alpha}} \sum_{\ell \geq 1} \ell^{k-1} a_\ell^\dagger A_\ell, \qquad
h^{(k)}= \sum_{\ell \geq 1} \ell^{k-1} \!\left[ A^\dagger_\ell A_\ell - \frac{1}{\alpha} a^\dagger_\ell a_\ell \right]\!
\end{gather*}
do satisfy the ${\widehat{\mathfrak s \mathfrak l}_2}$ commutation relations
\begin{gather*}
\big[e^{(k)},f^{(\ell)}\big]= h^{(k+\ell)} , \qquad \big[h^{(k)},e^{(\ell)}\big]= 2e^{(k+\ell)} , \qquad
\big[h^{(k)},f^{(\ell)}\big]= -2f^{(k+\ell)} .
\end{gather*}
 We thus get that
\begin{gather*}
 Q_1= \sqrt{\alpha} \big(f^{(0)}-e^{(0)}\big)\qquad\text{and}\qquad Q_2=h^{(1)}-h^{(0)}
\end{gather*}
and, as such, $\mathcal H_D$ is built from a special intertwining of $\hat {\mathcal H}_{y,z}^{(\alpha)}$ and $\hat {\mathcal H}^{(1)}_{y}$ with
the genera\-tors~$e^{(0)}$,~$f^{(0)}$ and $h^{(1)}$ of the nonnegative part
of ${\widehat{\mathfrak s \mathfrak l}_2}$.

This intertwining pattern is expected to hold for all the conserved quantities of the double CS model. Consider for instance the two conserved quantities of degree 1
\begin{gather*}
\mathcal I_D=(\alpha-1)\sum_{\ell \geq 1} A_\ell^+ A_\ell + \sum_{\ell\geq 1} \big[ a_\ell^\dagger A_\ell+
A_\ell^\dagger a_\ell \big],
\qquad
\mathcal P_D = \sum_{\ell \geq 1} \left[ A_\ell^+ A_\ell+ \frac{1}{\alpha}a_\ell^+ a_\ell \right],
\end{gather*}
 whose eigenvalues are respectively \eqref{ehat} and $|\lambda|+|\mu|$.
As for $\mathcal H_D$, the conserved quantity $\mathcal I_D$ can be written in terms of the usual conserved quantities
of the two CS models specif\/ied by $\hat {\mathcal H}_{y,z}^{(\alpha)}$ and $\hat {\mathcal H}^{(1)}_{y}$, and the (nonnegative-mode) generators of ${\widehat{\mathfrak s \mathfrak l}_2}$:
\begin{gather*}
\mathcal I_D = (\alpha-1)\mathcal P_y + [Q_1, \mathcal P_y],
\end{gather*}
where $\mathcal P_y$ is the momentum operator $\mathcal P$ in the variables $y_1,\dots,y_m$.
Similarly, we have
\begin{gather*}
\mathcal P_D = \mathcal P_y+\mathcal P_{y,z}.
\end{gather*}

\appendix

\section{The many combinations of ``super'', ``symmetric''\\ and ``polynomials''}\label{appendixA}

The supersymmetric CS model involves, in addition to the usual $N$ variables $x_i$, their anticommuting partners $\theta_i$. As a result, its eigenfunctions depend upon both set of variables. To describe a function with such a dependence on both types of variables, we use the terminology ``superfunction''. The qualif\/ier ``symmetric'', when applied to superpolynomials refers to the invariance upon the interchange of the pairs $(x_i,\theta_i)$ and $(x_j,\theta_j)$, an entanglement which is in the spirit of supersymmetry, as opposed to invariance under $x_i\leftrightarrow x_j$ and $\theta_k\leftrightarrow \theta_\ell$ separately. In this sense, it is thus natural to call the model's eigenfunctions the Jack superpolynomials.

Unfortunately, this terminology is a bit ambiguous with regard to existing similar nomenclatures. For instance, the symmetric superpolynomials just def\/ined are completely dif\/ferent
from the ``supersymmetric polynomials'' considered, e.g.,
 in~\cite{Stem} (which incidentally have nothing to do with the physicists' concept of supersymmetry and do not involve anticommuting variables). Indeed, the latter are def\/ined by two conditions: (1)~They are doubly symmetric polynomials in two
distinct sets of ordinary (commuting) variables $x_1,\dots, x_m$ and
$y_1,\dots, y_n$, i.e., invariant under
independent permutations of the $x_i$'s and the $y_i$'s.
(2)~They satisfy the following
cancelation condition: by substituting $x_1=t$ and $y_1=t$, the
polynomials become independent of $t$ (this is the meaning of ``supersymmetric'' in this context). An example of a generating
function for such polynomials is
\begin{gather*}
\prod_{i=1}^m (1-qx_i)\prod_{j=1}^n(1-qy_j)^{-1}= \sum_{r\geq 0}
p_{(r)}(x,y) q^r.
\end{gather*}

The terminology ``super-Jack polynomials'' is used by the authors of \cite{Ser,SV1,SV2} referring to objects also dif\/ferent from our Jack superpolynomials. Their construction originated from the realization that the CS model has an underlying $A_N$-type root structure. Viewed from this perspective, the model can be generalized to arbitrary root systems by still preserving integrability \cite{OP1,OP2}. The models considered in \cite{SV1,SV2} are natural extensions of this construction but based on Lie superalgebras (see also \cite{DH}). Here again, there are no anticommuting variables. By contrast, our Jack superpolynomials have no known Lie-algebraic interpretation.

Still another supersymmetric generalization of the CS model is presented in \cite{GK,KG}, albeit not formulated in terms of a supersymmetric quantum mechanical problem.
Here the Hamiltonian is induced from the radial part of the Laplace--Beltrami operator on symmetric superspaces. However, it does not contain anticommuting variables. The resulting models are a one-parameter generalizations of those of \cite{SV1,SV2}, but with the integrability issue unsettled.

To complete this list, we note that the term ``superpolynomial'' is used in the theory of knots, with a completely dif\/ferent meaning (it refers to the Poincar\'e polynomial for a quite general~-- triply-graded~-- homology)~\cite{knot}.

To avoid terminological confusion, we generally employ the nomenclature ``Jack polynomials in superspace'', using ``Jack superpolynomials'' as an abbreviation. But where is the superspace? Indeed, the usual variables $x_i$ do not refer to ordinary space.
However, the Jack polynomials are wave functions of a $N$-body quantum mechanical problem and as such they are def\/ined in the conf\/iguration space of the various particles described by the positions $x_i$, $i=1,\dots , N$. It is this conf\/iguration space that is lifted to superspace.

\subsection*{Acknowledgements}

We thank Olivier Blondeau-Fournier for his collaboration on~\cite{BLM}. This work was supported by the Natural Sciences and Engineering Research Council of Canada; the
Fondo Nacional de Desarrollo Cient\'{\i}f\/ico y
Tecnol\'ogico de Chile grant \#1130696.

\pdfbookmark[1]{References}{ref}
\LastPageEnding

\end{document}